\documentclass{JHEP3}
\usepackage{graphicx}
\usepackage{amsmath}
\usepackage{amsfonts}
\usepackage{amssymb}
\usepackage{epsfig}
\newcommand{\be}{\begin{equation}}
\newcommand{\ee}{\end{equation}}
\newcommand{\ben}{\begin{eqnarray}}
\newcommand{\een}{\end{eqnarray}}
\newcommand{\bb}{\bibitem}

\newcommand{\wt}{\widetilde}
\title{Regular and Periodic Tachyon Kinks}
\author{D. Bazeia, R. Menezes, and J.G. Ramos\\
Departamento de F\'\i sica, Universidade Federal da Para\'\i ba\\
Caixa Postal 5008, 58051-970 Jo\~ao Pessoa PB, Brazil}
\abstract{We search for regular tachyon kinks in an extended model,
which includes the tachyon action recently proposed to describe the tachyon
field. The extended model that we propose adds a new contribution to the
tachyon action, which allows obtaining stable tachyon kinks of regular profile,
which may appropriately lead to the singular kink found by Sen sometime ago.
Also, under specific conditions we may find periodic array of kink-antikink
configurations.\\

\vspace{1cm}
}
\maketitle
\begin{document}
\section{Introduction}

String theory is perhaps the most plausible candidate for a relativistic
theory to describe the electromagnetic, weak, strong, and gravitational
interactions altogether. It engenders a very rich structure
which includes stable or BPS and unstable or non-BPS branes. In the last
case, instabilities of non-BPS branes are marked by the presence of tachyon
fields, whose dynamics are directly related to the process in which
non-BPS branes decay into BPS branes --- see e.g.
Refs.~{\cite{t1,t2,t3,t4,t5,t6,t7,t8}}.

These recent investigations on tachyons have suggested
that the tachyon dynamics is described by the action
\be\label{a}
S=-\int d^{p+1}x\, V(T)\,
\sqrt{1+\eta^{\mu\nu}\partial_\mu T\partial_\nu T}
\ee
where $T=T({\vec x},t)$ is the tachyon field, real, and $x^\mu=(t,{\vec x})$
is the position vector. Also, the Minkowski metric has signature
$(-,+,+,+,\cdots)$, and $V(T)$ is the potential, which is non negative,
obeys $V(T\to\pm\infty)\to0$ and attains its global maximum at $T=0.$
We consider $0\leq V(T)\leq1.$

In the present work, in the above action we modify the Lagrange density
to the new form
\be\label{ma}
{\wt{\cal L}}=-V(T)\left(\sqrt{1+\eta^{\mu\nu}\partial_\mu T\partial_\nu T}-
\frac1{\sqrt{1+r^2F^2}}\right)
\ee
where $r$ is a parameter, real and positive, and
$F=F(T,\eta^{\mu\nu}\partial_\mu T\partial_\nu T).$
The real parameter $r$ is introduced to control the way one goes
beyond former investigations: for $F$ that behaves appropriately,
the limit $r\to\infty$ gives the former action (\ref{a}) and leads to
the problem investigated by Sen in Ref.~{\cite{S}};
for $r>>1$ one may get to the recent investigation \cite{css}, in which
one modifies the action (\ref{a}) by adding a term which depends on the
derivative of the tachyon field. In Ref.~{\cite{css}}, the term included in the
action is controlled by a very small parameter, and admits a
perturbative investigation which very nicely leads to regular kink, and gives
the singular kink of Ref.~{\cite{S}} in the appropriate limit. 

In this work, in the modification we are introducing in (\ref{ma}) we consider
functions that only depend on the tachyon field, that is, we
consider $F=F(T)$ as a non negative function $T$ alone. Moreover,
we suppose that $F(T)$ is limited to some interval, that is, we suppose
that $0\leq F(T)\leq 1.$ As we shall show, the above model engenders specific
features which are of direct interest to high energy physics. To do this,
we organize the subject of this work in the sections that follow, where we
study the presence of tachyon kinks of regular and periodic profile.

\section{Generalities}

We use the modified model to examine the energy corresponding to static
configuration $T=T({\vec x})$. We get
\be
E=\int d^px\,V(T)\left(\sqrt{1+\partial_iT\partial^iT}-
\frac1{\sqrt{1+r^2F^2}}\right)
\ee
We investigate stability of static solutions with the Derrick-Hobart theorem
\cite{H,D,J,bmm} --- see also Ref.~{\cite{css}}. We change
$T({\vec x})\to T^\lambda({\vec x})=T(\lambda{\vec x})$ to get to
the condition for stability of the solution $T({\vec x})$
\be
\frac{p+(p-1)\partial_i T\partial^i T}{\sqrt{1+\partial_i T\partial^i T}}=
\frac{p}{\sqrt{1+r^2F^2}}
\ee
The case of a single spatial dimension is special. Here the above condition
reduces to the simpler form $T^{\prime 2}(x)=r^2 F^2(T),$ or better
\be\label{bps}
T^{\prime}(x)=\pm r F(T)
\ee
where the prime stands for derivative with respect to $x.$ We note
that the above Eq.~(\ref{bps}) reproduces first-order differential
equations that appear in the bosonic sector of supersymmetric field
theory described a single chiral superfield; see, e.g., Ref.~{\cite{bb00}}.

Let us now investigate the equation of motion that follows from the
modified action. It can be written in the form
\be
\frac1{\sqrt{1+\partial_\mu T\partial^\mu T}}
\left(\frac{dV}{dT}-V\partial_\mu\partial^\mu T+
V\frac{\partial^\mu T\partial^\nu T\partial_\mu\partial_\nu T}
{1+\partial_\mu T\partial^\mu T}\right)=\frac1{\sqrt{1+r^2F^2}}
\left(\frac{dV}{dT}-\frac{r^2 F V}{1+r^2F^2}\frac{dF}{dT}\right)
\ee
In the case $p=1,$ the static field obeys
\be\label{em1}
\frac1{T^\prime}\frac{d}{dx}\left(\frac{V}{
\sqrt{1+T^{\prime2}}}-\frac{V}{\sqrt{1+r^2F^2}}\right)=0
\ee  
This equation is solved by $T^\prime\to\infty,$ which are stable solutions
of the equation (\ref{bps}) in the limit $r\to\infty$, which leads to the
case first investigated by Sen in \cite{S}, giving rise to the stable
but singular tachyon kinks
\be\label{sk}
T^{\pm}_S(x)=\begin{cases}
\,\pm\infty\,& {\rm for}\,\,\, x>0
\\ \,0\,& {\rm for}\,\,\, x=0
\\ \,\mp\infty\,& {\rm for}\,\,\, x<0
\end{cases}
\ee
There are other solutions, which obey
\be\label{s}
T^{\prime 2}=\frac{1}{\left(\frac{V_0}{V}+
\frac{1}{\sqrt{1+r^2F^2}}\right)^2}-1
\ee
where $V_0$ is a real constant, constrained to obey
\be\label{cons}
0\leq{\frac1{\sqrt{1+r^2F^2}}+\frac{V_0}{V}}\leq1
\ee
The case $V_0=0$ reproduces the former
Eq.~(\ref{bps}), leading to the conclusion that the solutions of the
equation of motion are stable for $V_0=0.$

In our model, in the case $p=1$ the unidimensional stable static solutions
have energy in the form
\be\label{e}
E=r^2\int dx \frac{V(T)F^2(T)}{\sqrt{1+r^2F^2(T)}}
\ee
In the case $r\to\infty$ we get
\be\label{es}
E_S=\int_{-\infty}^{\infty} dT\, V(T)
\ee
which requires that the tachyonic potential be integrable. Another case
is given by $r=0.$ Stable solutions should satisfy $T^\prime=0,$ which
makes the tachyonic field constant. The energy associated to such constant
configurations vanishes. The energy in (\ref{e}) is non-negative, and for
stable solutions it varies in the interval $0\leq E\leq E_S,$ for functions
$F(T)$ which behave appropriately.

We notice that in the modified model, in the case $p=1$ stable tachyon
solutions obey $T^\prime=r F(T)$, and so they do not depend on the explicit
form of the tachyon potential. However, the tachyon potential plays the
important role of controlling the energy of the tachyon solution. This fact
helps us to understand why the singular tachyons (\ref{sk}) are stable,
finite energy solutions.

We also notice that in the modified model, the energy-momentum tensor
$T_{\mu\nu}$ gives the energy density $T_{00}$
\be
T_{00}\equiv V(T)\left(\sqrt{1+T^{\prime 2}}-\frac1{\sqrt{1+r^2F^2}}\right)
\ee
and pressure along the non trivial $x$ direction $(P_1=T_{11})$
\be\label{t11}
T_{11}\equiv -\frac{V}{\sqrt{1+T^{\prime 2}}}+\frac{V}{\sqrt{1+r^2F^2}}
\ee
The pressure is constant since $T_{11}^{\prime}(x)=0$ --- see Eq.~(\ref{em1}).
We see that $T_{11}=-V_0,$ where $V_0$ is the constant that we have introduced
to write Eq.~(\ref{s}). Thus, the case $V_0=0$ corresponds to vanishing
pressure, and gives rise to stable finite energy tachyon configurations
which obey $T^{\prime}=\pm r F.$ Furthermore, since $V_0$ must obey the
constraint (\ref{cons}), we can also have two other distinct possibilities:
one for $V_0$ positive, representing the case of negative pressure,
and the other for $V_0$ negative, representing the case of positive
pressure. We shall show below that in the case of non vanishing $V_0,$
we must compactify the real line in order to have finite energy,
and this will give rise to periodic tachyon kinks. Thus, in the modified
model we shall find stable and regular tachyon kinks in an environment with
vanishing pressure. And also, we shall find periodic kink-antikink array
in another environment, with negative pressure. We notice
from the above Eq.~(\ref{t11}) that the limit $r\to\infty,$ which leads to
the standard tachyon action, gives rise to the case of negative pressure,
and nothing more --- see Ref.~{\cite{kkl,bms}}.

\section{Specific models}

We investigate the case with $V_0=0,$ which corresponds to vanishing pressure.
This case gives rise to models that support stable tachyon kinks of regular
profile. The energy depends on the tachyon potential, $V(T)$. Thus, we choose
the tachyon potential such that $E_S=1.$ In this case the energy of the tachyon
configurations is restricted to be in the interval $0<E<1$. We may choose
$V_{I}(T)=\exp(-\pi T^2),$ $V_{IIa}(T)=1/2\cosh^2(T)$,
$V_{IIb}(T)=1/\pi\cosh(T),$ and $V_{III}(T)=1/\pi(1+T^2)$
which identify type-I, type-II, and type-III models, all leading
to unit energy. We will study these models to have a better understanding
of the role of the potential for the modified tachyon action that we propose
in (\ref{ma}). Specific models involving the choices $F(T)=1$ and
$F(T)=1/\cosh(T)$ will be investigated below.

\subsection{Type-I models}

We consider the case $F(T)=1,$ which
implies that $T^\prime=\pm r$, giving rise to the solutions $T_\pm(x)=\pm r x,$
which lead to the singular kinks (\ref{sk}) in the limit $r\to\infty.$
This case reproduces the solutions of Ref.~{\cite{css}}. The energy
corresponding to these solutions has the form
\be
E_I(r)=\frac{r^2}{\sqrt{1+r^2}}\int_{-\infty}^{\infty} dx \,V(T)
\ee
Thus, for $V(T)=\exp(-\pi T^2)$ and for $T_\pm=\pm r x$ we get
\be\label{e1}
E_I(r)=\frac{r}{\sqrt{1+r^2}}
\ee
We use this result to get $E(r=0)=0.$ This is interesting, since the limit
$r\to0$ leads to a constant tachyon configuration. Our model gives vanishing
energy for trivial constant tachyon configurations at $r=0,$ and unit energy
for the singular kink (\ref{sk}) at the limit $r\to\infty.$ The
stable kink solutions $T^{\pm}(x)=\pm r x$ are parametrized by $r$,
and have energy as in Eq.~(\ref{e1}), which is well-defined in the entire
interval $0\leq r\leq1.$ In Fig.~[1] we plot $E_I(r)$ in the whole interval
$r\in[0,\infty).$

We also consider the case of $F(T)=1/\cosh(T)$ to get
\be\label{rk}
T^\prime=\pm \frac{r}{\cosh(T)}
\ee
This equation was already solved in Ref.~{\cite{b}}. The solutions are
$T(x)=\pm\,{\rm arcsinh}(r\,x)$, and we realize that the singular
kink (\ref{sk}) is now very naturally recovered in the limit $r\to\infty.$
The energy of the regular kinks can be written as
\be
{\wt E}_I(r)=r\,\int_{-\infty}^{\infty}dx\,
\frac{e^{-\pi {\rm arcsinh}^2(x)}}{(1+x^2)\sqrt{1+\frac{r^2}{1+x^2}}}
\ee
and depends on the parameter $r.$ It vanishes for $r=0$, and converges
to unit in the limit $r\to\infty.$ In Fig.~[1] we plot the energy density
for $r\in[0,\infty).$ We note that ${\wt E}_I(r)$ is very close to $E_I(r),$
indicating that the choice of $F(T)$ determine no qualitative behavior. 

\subsection{Type-II models}

We first consider type-IIa models, and use $F(T)=1.$ The investigation is
similar to the former case. The kink solutions and the energy give the very
same results already obtained in the corresponding type-I model. Thus,
we consider the next case: $F(T)=1/\cosh(T).$ Here the energy changes to 
\be
{\wt E}_{IIa}(r)=\frac{r}{2}\,\int_{-\infty}^{\infty}dx\,
\frac{1}{(1+x^2)^2}\frac{1}{\sqrt{1+\frac{r^2}{1+x^2}}}
\ee
It vanishes for $r=0$, and converges to unit in the limit $r\to\infty.$
In Fig.~[1] we plot the energy ${\wt E}_{IIa}(r)$ in the entire interval
$r\in[0,\infty).$ We note that ${\wt E}_{IIa}(r)$ is very close
to ${\wt E}_{I}(r),$ suggesting that the specific choice of the tachyonic
potential seems to determine no qualitative behavior.

We also consider type-IIb models, with the potential $1/\pi\cosh(T).$ 
For $F(T)=1$ we get the very same result already obtained in the
former case. For $F(T)=1/\cosh(T),$ the investigation is slightly modified,
with the energy changing to
\be
{\wt E}_{IIb}(r)=\frac{r}{\pi}\,\int_{-\infty}^{\infty}dx\,
\frac{1}{(1+x^2)^{3/2}}\frac{1}{\sqrt{1+\frac{r^2}{1+x^2}}}
\ee
In Fig.~[1] we also plot ${\wt E}_{IIb}$ as a function of $r.$ We see that
it is similar to ${\wt E}_{IIa},$ showing that tachyon potentials of the form
$1/\cosh(T)$ and $1/\cosh^2(T)$ give very similar results.

\subsection{Type-III models}

Again, we first consider the case $F(T)=1.$ The investigation is similar
to the former cases, and both the kink solutions and the energy give the
very same results already obtained, Thus, we consider the next
case of $F(T)=1/\cosh(T).$ The kink solutions are the same, but the energy
changes to 
\be
{\wt E}_{III}(r)=\frac{r}{\pi}\,\int_{-\infty}^{\infty}dx\,
\frac{[1+{\rm arcsinh}^2(x)]^{-1}}{(1+x^2)\sqrt{1+\frac{r^2}{1+x^2}}}
\ee
It vanishes for $r=0$, and converges to unit in the limit $r\to\infty.$
However, the convergence is very slow, due to the specific form of the
tachyon potential in this case.
In Fig.~[1] we also plot ${\wt E}_{III}$ as a function of $r.$ Its behavior
is now distinct from the others, and this shows that the inverse power
behavior of the tachyon potential leads to distinct energy behavior
for tachyon kink, for $F(T)\neq1$.

\begin{figure}[h]
\includegraphics[{height=8.0cm,width=12.0cm}]{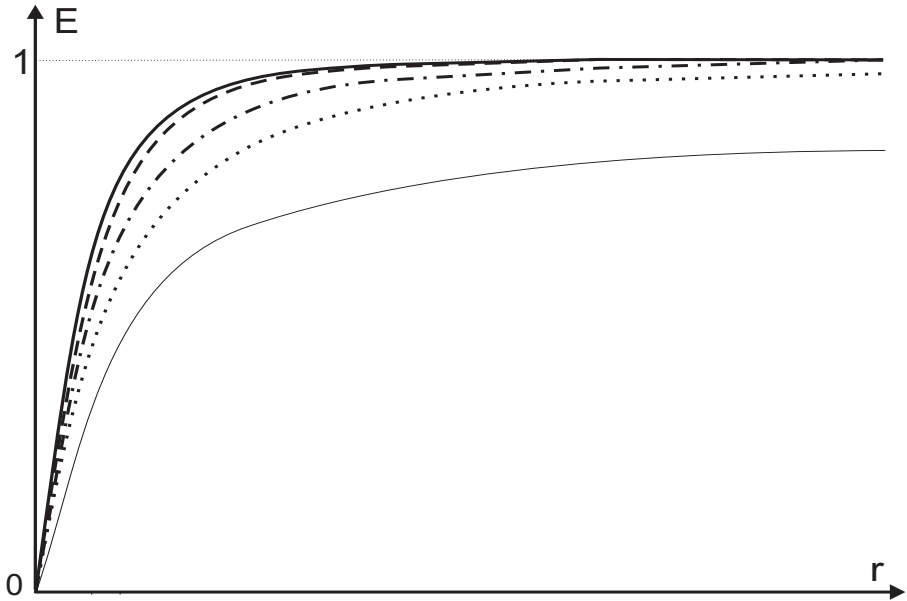}
\caption{The energy as a function of the real parameter $r.$ The thick line
corresponds to $F(T)=1,$ for all the models. The dashed, dash-dotted,
and dotted lines correspond to $F(T)=1/\cosh(T)$ for the type-I, type-IIa,
and type-IIb models, respectively. The thin line corresponds to the type-III
model, for $F(T)=1/\cosh(T)$.}
\end{figure}

\section{Periodic solutions}

We now consider other possibilities, which appear when $V_0\ne0.$ As we have
already seen, for $V_0\neq0$ there are no stable solutions. However, we can
compactify the real line to investigate static, periodic and finite energy
solutions. We follow Refs.~{\cite{kkl,bms}}, which has
already investigated the presence of static, periodic and finite energy
solutions in the model (\ref{a}).

In order to study static, periodic and finite energy solutions in the modified
model, we consider the case $V(T)=V_{IIa}(T)=1/2\cosh^2(T),$ and the case
$F(T)=1.$ We get
\be
T_p^{\prime\,2}(x)=
\frac1{[1/\sqrt{1+r^2}+2V_0\cosh^2(T)]^2}-1
\ee
The constant $V_0$ is now restricted to belong to an interval which depends
on $r.$ A specific case is $r=0,$ and now we have $-1/2\leq V_0\leq0,$ with the
tachyon field constrained to live in the interval $-T_0\leq T\leq T_0,$ with
$T_0={\rm arccosh}\sqrt{-1/2V_0}.$ This case corresponds to positive pressure.

We use $r=0$ to define the function
\be
G(T)=\frac1{[1+2V_0\cosh^2(T)]^2}-1
\ee
In this case, we notice that the points $\pm T_0$ are divergences for $G(T),$
and so the periodic solutions should end vertically, with divergent
derivative. This behavior is not admissible. Indeed, there are no finite energy
periodic solutions for $V_0<0,$ in the case of positive pressure. 
We see this using the energy density to write, for $F(T)=1,$ and for $r=0:$
$T_{00}(x)=-V_0/(1+V_0/V).$ We take $V=V_{IIa}$ to get
\be\label{ed2}
T_{00}(x)=-\frac{V_0}{1+2V_0\cosh^2(T)}
\ee
The constant $V_0$ should obey $-1/2\leq V_0\leq0,$ and the energy density
diverges at the values $\pm{\rm arccosh}\sqrt{-1/2V_0}.$ By the way, we notice
that the energy vanishes in the limit $V_0\to0,$ because the choice
$F(T)=1$ gives rise to tachyon kinks which obey $T^{\prime}(x)=\pm r,$ and for
$r=0$ we get to trivial tachyon solutions with vanishing energy.

We now consider the case $r\to\infty,$ which gives $0\leq2V_0\cosh^2(T)\leq1.$
There are solutions for $0\leq V_0\leq1/2,$ with
$-{\rm arccosh(\sqrt{1/2V_0})}\leq T\leq{\rm arccosh(\sqrt{1/2V_0})}.$
This case corresponds to negative pressure, and the solutions are similar
to the periodic kinks found in Ref.~{\cite{kkl,bms}}.
In the modified model, $V_0$ varies in the interval
$[-1/2\sqrt{1+r^2},1/2-1/2\sqrt{1+r^2}],$ which depends on $r.$ For $V_0=0,$
there are finite energy tachyon kinks which engender very nice profile, and
for $V_0$ positive there are periodic solutions similar to the solutions
found in Ref.~{\cite{kkl,bms}}.

We can illustrate both cases involving positive and negative pressure with
simpler models. We consider the case of a single real scalar field to write
the Lagrange density
\be
{\cal L}= -\frac12\partial_\mu\phi\partial^\mu\phi-
\frac12\left(\frac{dW}{d\phi}\right)^2
\ee
where $W=W(\phi)$ is a smooth function of $\phi.$ The case of periodic
tachyon kinks for negative pressure is similar to the case described by
\be
W(\phi)=\frac12\phi\sqrt{1-\phi^2}+\frac12 {\rm arcsin(\phi)}
\ee
which requires that $\phi\in [-1,1].$ In this case the first-order equations
are given by
\be
\frac{d\phi}{dx}=\pm\sqrt{1-\phi^2\,}
\ee
They are solved by $\phi(x)=\pm\sin(x)$ and this requires that
$x\in [-\pi/2,\pi/2].$

We can make an array of kink-antikink by alternating kinks and antikinks in
the real line. The energy density of each kink or antikink is
$\varepsilon(x)=\cos^2(x),$ which can be integrated in the interval
$-\pi/2\leq x\leq\pi/2$ to give $E_k=\pi/2.$ Thus, for $N$ kink-antikink
pairs we get $E_k^N=N\pi,$ and the size $L_k^N$ of the array has to obey
$L_k^N=2N\pi.$

The above kink or antikink is stable. We write
$\phi(x,t)=\phi(x)+\sum_n\,\eta_n(x)\,\cos(w_nt)$ in order to obtain the
Schr\"odinger-like equation for stability
\be
-\frac{d^2\eta_n}{dx^2}-\eta_n=w^2\,\eta_n
\ee
The energies are such that $w_n=\sqrt{n(n+2)\,},$ and the eigenfunctions are
even or odd, for $n$ even or odd, respectively. They are given by
\be
\eta_n(x)=\sqrt{2/\pi}\,
\begin{cases}
\cos[(n+1)x]\;\;\;{\rm for}\;n=0,2,4,...
\\ \\
\sin[(n+1)x]\;\;\;{\rm for}\;n=1,3,5,...
\end{cases}
\ee
The gap $w_{n+1}-w_n$ starts at $\sqrt{3}$, and converges to $1$ for
increasing $n.$ This spectrum is similar to the spectrum found in
Ref.~{\cite{mz}}.

The case of positive pressure is different. This phase is unstable,
and the energy of kink or antikink diverges. We illustrate this situation
with the model
\be
V(\phi)=\frac12 {\rm sec}^2(\phi)
\ee
which is described by $W(\phi)=\ln[{\rm sec}(\phi)+\tan(\phi)]$ and requires
that $\phi\in [-\pi/2,\pi/2].$ In this case the first-order equations are
\be
\frac{d\phi}{dx}=\pm {\rm sec}(\phi)
\ee
These equations are solved by $\phi(x)=\pm\arcsin(x)$ for $x\in[-1,1].$ The
derivative diverges for $x\to\pm1.$ Also, the energy density is
$\varepsilon(x)=1/(1-x^2),$ which diverges in the limit $x\to\pm1,$
leading to divergent energy. We also notice that $1/\sqrt{1-x^2\,}$ would
be the zero mode, but this is not normalizable in the interval $x\in[-1,1].$

\section{Ending comments}

In this work we have modified the tachyon action (\ref{a}) by changing
the square root contribution as in (\ref{ma}). We have found stable, finite
energy tachyon kinks in the case of vanishing pressure in several different
models. We have also found a network of kink-antikink configurations in the
case of negative pressure. Although the scenario one finds in the case of
negative pressure is similar to other extensions that appeared recently,
the case of vanishing pressure is new, and it supports regular, stable
and finite energy tachyon kinks which nicely lend thenselves to Sen's
singular solutions in the appropriate limit.   

The modification that we introduce gives rise to another phase, corresponding
to the case of positive pressure. However, the phase with positive pressure is
unstable, giving rise to tachyon kink and antikink which end with divergent
derivative, signalling the presence of divergent energy. The cases concerning
positive and negative pressure have interesting analogies with field-theoretic
models, which allowed illustrating both possibilities within simpler scenarios.

We thank F.A. Brito for discussions, and CAPES, CNPq, PROCAD/CAPES
and PRONEX/FAPESQ/CNPq for partial support.


\end{document}